\documentclass[conference]{IEEEtran}
\IEEEoverridecommandlockouts
\ifCLASSINFOpdf
  \usepackage[pdftex]{graphicx}
\else
\fi
\begin{document}
%
\title{Leveraging Team Dynamics to Predict Open-source Software Projects’ Susceptibility to Social Engineering Attacks\\


\thanks{Authors are listed alphabetized by last name.}
\thanks{This material is based upon work supported by (while serving at) the National Science Foundation (NSF). Giovanini and Oliveira's contributions were supported by NSF CNS-1513572.}
}

\author{
\IEEEauthorblockN{Luiz Giovanini}
\IEEEauthorblockA{\textit{Dept. of Electrical and} \\\textit{Computer Engineering} \\
\textit{University of Florida}\\
Gainesville, USA \\
lfrancogiovanini@ufl.edu}
\and
\IEEEauthorblockN{Daniela Oliveira}
\IEEEauthorblockA{\textit{Dept. of Electrical and} \\\textit{Computer Engineering} \\
\textit{University of Florida}\\
Gainesville, USA \\
daniela@ece.ufl.edu}
\and
\IEEEauthorblockN{Huascar Sanchez}
\IEEEauthorblockA{\textit{Computer Science Laboratory}\\
\textit{SRI International}\\
Menlo Park, USA \\
huascar.sanchez@sri.com}
\and
\IEEEauthorblockN{Deborah Shands}
\IEEEauthorblockA{\textit{Computer Science Laboratory}\\
\textit{SRI International}\\
Menlo Park, USA \\
deborah.shands@sri.com}
}

\newcommand{\tool}{\textsf{ToolName}}
\newcommand{\SE}{SE}
\newcommand{\XXX}{\textbf{***CHECK***}}

\maketitle

\begin{abstract}
Open-source software (OSS) is a critical part of the software supply chain.  Recent social engineering attacks against OSS development teams have enabled attackers to become code contributors and later inject malicious code or vulnerabilities into the project with the goal of compromising dependent software. The attackers have exploited interactions among development team members and the social dynamics of team behavior to enable their attacks. We introduce a security approach that leverages signatures and patterns of team dynamics to predict the susceptibility of a software development team to social engineering attacks that enable access to the OSS project code. The proposed approach  is programming language-, platform-, and vulnerability-agnostic because it assesses the artifacts of OSS team interactions, rather than OSS code.
\end{abstract}


%
\IEEEpeerreviewmaketitle

\section{Introduction}

Today's software supply chain depends heavily on externally written open-source software (OSS) made available through packages, libraries, or modules~\cite{synopsisReport19}.  
The OSS paradigm has facilitated and catalyzed social engineering (SE) attacks because of the openness of the  ecosystem  and the possibility that anyone can become a contributor. This emerging and dangerous  threat to the  software supply chain was exemplified in the 2018 attack against the {\tt npm}  package {\tt event-stream}~\cite{constantin2018}. In this case, the attacker (a code contributor) became the maintainer of the {\tt event-stream} package by convincing the
original code maintainer to transfer {\tt event-stream}'s ownership.  The attacker then created a dependency to malicious code that ran only in
environments operating a Bitcoin wallet.  The malicious code was designed to steal users' private cryptographic keys and send them to an attacker-controlled
server, enabling the attacker to later steal Bitcoins from end users. This attack against 
involved multiple stages
of software dependencies and a sequence of \SE\ attacks against the original code maintainer, producing a
deeply obfuscated back door. 

\begin{figure*}[t]\centering
  \frame{\includegraphics[width=0.75\linewidth]{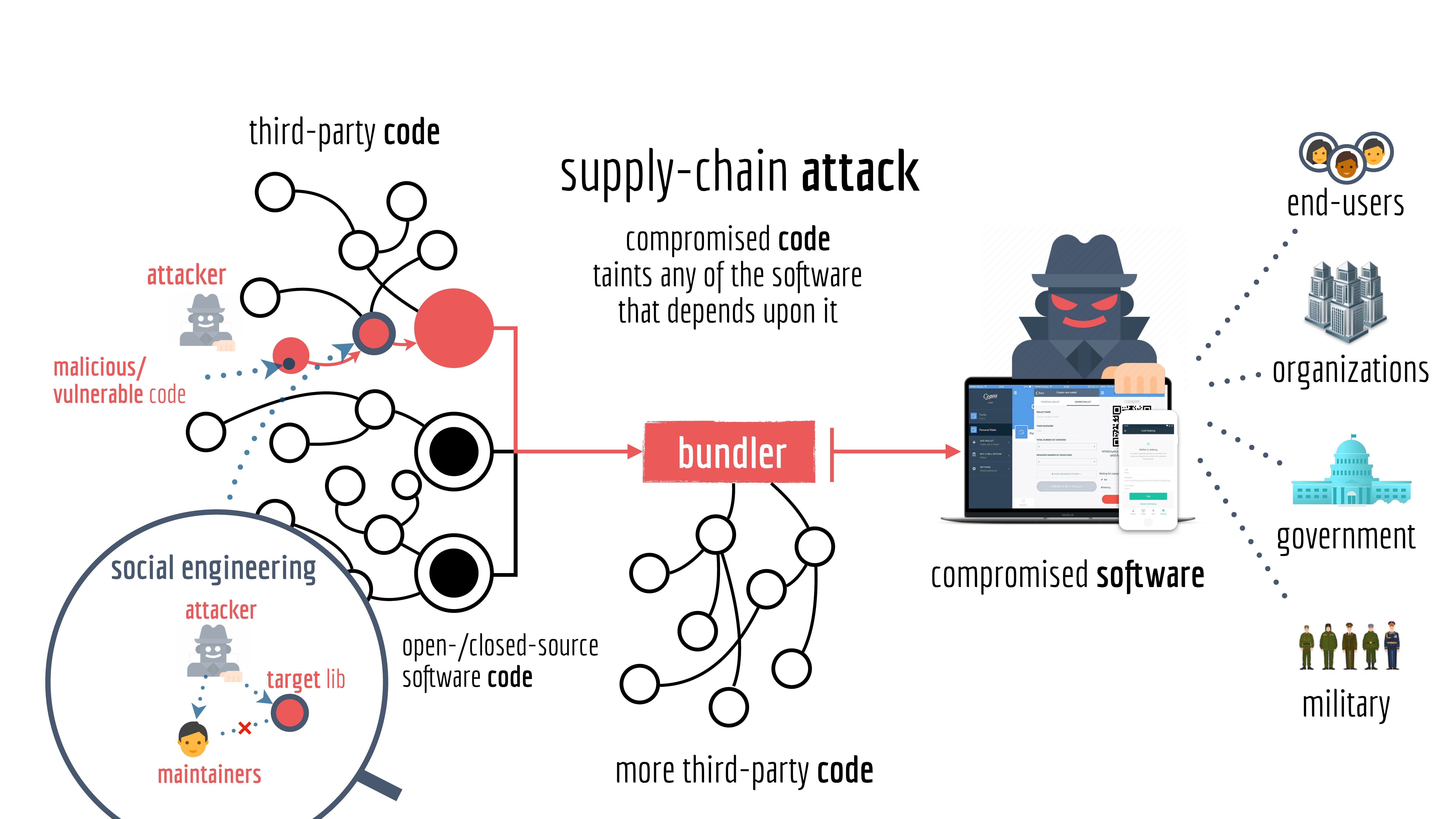}}
 \caption{A \SE-based supply-chain attack against an OSS ecosystem.}
  \label{fig:supply-chain-attach}
\end{figure*}

Figure \ref{fig:supply-chain-attach} illustrates a generalized view of this \SE\ attack on supply-chain OSS. In the figure, an attacker uses \SE\ tactics to become a contributor to a popular OSS project and injects malicious code or software vulnerabilities into the code base. Later, developers of other open- or closed-source software make use of the compromised OSS as a component or a dependency. An attack that compromises popular OSS code is dangerously scalable, as it can target many users in a variety of sectors and types of organizations (corporate, government, military).  For example, from September 2018 to 2019, the {\tt event-stream} package was downloaded almost 80,000 times. Research on supply-chain attacks on OSS ecosystems is still in its infancy~\cite{zimmermann2019small, duan2020measuring} and focuses on the technical aspects of the compromise, such as detection of suspicious or
typo-squatted packages in package managers. To the best of our knowledge, no work has attempted to address the \SE\ component of such attacks from a
socio-technical angle.

In this paper, we introduce a novel security approach that focuses on the relationship between team dynamics and OSS project susceptibility to \SE\
attacks. \textit{Team dynamics} emerge from common, ongoing interactions among one or many development team members as they design, vet, and implement OSS.
Team dynamics have signatures that can be identified from signals such as: team discussions, consumption of application programming interfaces (APIs) and
service provider interfaces (SPIs) and adoption of programmatic idioms, lightweight code review workflows, adherence to coding conventions, maintenance
and software evolution, and products (e.g., commits, technical debt, bugs, and vulnerabilities).

Details of the {\tt event-stream} attack and other attacks against OSS projects suggest that fragile team dynamics may be an important enabler of \SE\ attacks
against OSS development teams. \textit{Fragile team dynamics} are behavioral signals (e.g., maintainer burnout) and interactions (e.g., controversial or
toxic communication, maintenance frequency, frequent merge request rejections, autocratic decisions, classism) among development team members that an attacker
may exploit to gain access to the project. As demonstrated in the {\tt event-stream} attack, by looking at how frequently this package was maintained, the attacker was able to detect the maintainer burnout signal, offer help, and acquire the permissions needed to
execute the attack.

If  fragile team dynamics correlate with the susceptibility of teams to \SE\ attacks targeting the software supply chain, then observations of these dynamics
and their evolution in OSS projects could enable next-generation, learning-based \SE\ attack prediction that is programming language-, platform-, and
vulnerability-agnostic. Automatically identifying fragile team dynamics may enable us to predict which OSS projects are more likely to fall victim to \SE\
attacks in the future.  Open and closed-source developers could use this information to avoid creating risky dependencies on code created or maintained by teams with fragile dynamics.  These predictions could also enable teams to self-diagnose team dynamics that may be slipping into fragile behaviors.

As we consider the relationship between team dynamics and susceptibility of OSS projects to \SE\ attacks, key research questions arise, including: over time, do teams experience shifts in their susceptibility to the deception and influence that enable \SE\ attacks? What causes such shifts and can they be detected? Are there patterns of team dynamics that correlate with the introduction of security vulnerabilities in OSS projects, even if those vulnerabilities are later patched? How can we distinguish unproductive team dynamics, which might result in poor-quality code and lead to project abandonment, from fragile team dynamics that preserve acceptable code quality, but might allow unsafe/malicious code to be added to the project? What metrics of fragility in team dynamics could be useful as predictors of susceptibility to \SE\ attacks?

 
 

  



The remainder of this paper describes the vulnerability of the software supply chain to \SE\ attacks (Section~\ref{sec:supplychainvuln}), a data-driven approach to prediction using signals and patterns of team dynamics (Section~\ref{sec:datadrivenapproach}), related works 
(Section~\ref{sec:relatedwork}), and our vision for a security approach that predicts future susceptibility of teams to \SE\ attacks rather than focusing on detecting present vulnerabilities (Section~\ref{sec:conclusion}).

\section{Vulnerability of the software supply chain to social engineering}
\label{sec:supplychainvuln}






In the past, attackers relied on finding latent vulnerabilities in code
executing in the victim’s system. However, given that the  most widely used software today is heavily dependent on OSS (e.g., software modules from the Web, code repositories, etc.), this ecosystem is now the new
frontier for exploitation by adversaries. 
An emerging, dangerous trend is that adversaries plant software vulnerabilities and malware in OSS code likely to be deployed by the victim, thus creating a ``poison pill'' that the victim is likely to swallow.  The attacker can later target other victims using the planted vulnerabilities and malware. One way to facilitate such latent attacks is \SE, which is increasing in OSS ecosystems~\cite{constantin2018}. For example, one \SE-based software supply chain attack from 2018 targeted 18 companies and affected 2.3 million computers~\cite{CCleaner2018}.

Although \SE\ attacks differ from one another, Salahdine et
al.~\cite{salahdine2019social} indicate that they have a common pattern and involve similar stages. First, the adversary collects information about the target (e.g., infrequent maintenance of the code base, blindspots in their APIs).
Second, the adversary builds trust by developing a relationship with legitimate maintainers and contributors of the project to be perceived as a real and trusted maintainer.
Third, the adversary exploits the available information and
executes the attack. 
The sophisticated, multifaceted, socio-technical nature of such
an attack challenges conventional detection methods~\cite{krombholz2015advanced,breda2017social}.



Unfortunately, measuring and detecting \SE\ attacks targeting the software supply chain remains a challenge~\cite{duan2020measuring}. While static and dynamic analysis tools can help OSS maintainers to assess software safety~\cite{henley2018cfar, garrett2019detecting}, many attacks start with members of the OSS project; that is, maintainers, trusted contributors, and other developers who may not believe they are vulnerable to \SE\ attacks. Without an understanding of the complex team dynamics (and social interactions) that constantly shape an OSS project, existing solutions can  only do so much to identify contexts most likely to elicit a \SE\ threat.

Interpersonal interactions and collaboration are key ingredients of teams
thriving in organizations \cite{kabo2018architecture}. 
Social engineers often rely on socio-psychological aspects and relate to group dynamics and interaction to develop a relationship that will increase their chances of success with their future victims~\cite{ krombholz2015advanced,albladi2020predicting}. 
Social engineering is highly effective because it relies on human cognitive biases and susceptibility to influence, rather than
vulnerabilities in the software supply chain. Mistakes made by members of
an OSS project are much less predictable than malware-based intrusions~\cite{nagrath2016protocols}, making them more difficult to identify and prevent via purely
technical protection mitigation techniques. Humans are highly susceptible to deception~\cite{albladi2020predicting} and thus may be
easily lured into making security mistakes or giving away sensitive
information~\cite{salahdine2019social}.

\section{A data-driven approach to prediction}
\label{sec:datadrivenapproach}

In this section, we describe our framework for leveraging team dynamics to predict OSS project susceptibility to \SE\ attacks in a data-driven manner. 
Our main challenges lie in choosing which data to inspect and which to ignore, identifying the features of the data that matter, distinguishing patterns of team dynamics that may have different impacts on code, and providing actionable metrics for predicting susceptibility to future \SE\ attacks.
We also expect to encounter somewhat generic challenges (e.g., in data acquisition and storage infrastructure, data pre-processing and filtering), but we will not discuss them here unless they have special bearing on the team dynamics analysis.

\subsection{Data collection}
\label{subsec:datacollect}

Communications among team members, as well as logs of project-related activities are written and publicly available. We can use tools to extract data from OSS projects hosted on a platform like GitHub (one of the largest and most popular open-source code hosting sites in the world) and analyze it to better understand characteristics of project team dynamics. Therefore, there is no shortage of available data.

\subsection{Modeling interactions among software developers}
\label{subsec:tempt}


Consider the problem of predicting the susceptibility of a group of software developers to \SE\ attacks targeting the software supply chain. OSS  developers interact by \textit{collaboratively evolving an OSS}: they discuss, create, or update software artifacts critical to the OSS. These interactions have clear OSS outcomes (e.g., creating a new dependency, a new API, a broken build). We can naturally represent software developers and project outcomes (as nodes) and their actions (as edges) using a graph. An interaction (or an action instance) between software developers occurs when these  developers discuss, create, review, and edit a software artifact in a short succession, and
one of them either approves, rejects, or undoes, the work of the other (i.e., a project outcome). 

\sloppy{
Software developer interactions are dynamic because contributors, activities, and interests change over time. To account for these temporal changes, we model them as time-dependent interactions. 
We can use graphs to help formulate the problem of \textit{team susceptibility to social engineering-based supply-chain attacks (TEMPT)} as a sequential or temporal learning problem over graphs.}
For example, we can model the {\tt event-stream} attack using a bipartite graph. Each edge in the graph is labeled with a feature vector at each time -- e.g., (1) general features including maintainer burnout level, presence of influence cues messages exchanged between the maintainer and the attacker, rate of interaction, gaps between consecutive exchanges, (2) action-specific features such as commenting, liking, ownership transferring, emotional salience of message (positive or negative), direction information, target artifact, etc. A single vector captures all of these features via an aggregation operation. Each vertex in the graph is associated with a vector of developer-specific features that might represent information about popularity, years of project involvement, position in the community (e.g., rockstar, trusted developer), active years on GitHub, technical skills, number of followers, contribution frequency, commits to failures ratio, reactions, etc. By capturing these time-dependent interactions among developers, the model will support analysis of teams and individuals.



\subsection{Feature learning for predicting social engineering-based
  supply-chain attack based on team dynamics}
\label{subsec:flearn}

Given the vast amount of software development data on code-hosting sites like GitHub, one of the major challenges for the proposed framework is feature engineering 
(a laborious and manual process that often requires some domain expertise).
We can focus on automated approaches to learning meaningful representations to support our multiple predictive tasks. Out of the box, the application of representation learning algorithms such as temporal word embeddings \cite{mikolov2013distributed} 
may be infeasible. However, this will give us the opportunity to tailor new feature learning approaches to the \SE\ prediction challenge.


\subsection{Team dynamics signatures to predict susceptibility to social engineering attacks} 
\label{subsec:tdynamics}
A key ingredient in our vision is the identification of team dynamics signatures that predict susceptibility to \SE-based supply chain attacks.
Identifying these signatures is like looking for needles in a haystack. For example, although team discussions seem to be an interesting team signature, they might not be a useful predictor of susceptibility because certain patterns of communication considered unpleasant
to other communities may be common among OSS developers \cite{kidane2007correlating} (e.g., bad communication style, lack of joy or positiveness). 


A sociological perspective 
we can use to study 
team dynamics is Dramaturgy. The dramaturgical analysis method~\cite{kivisto2007goffman} provides tools for examining individuals within the context of a social whole (e.g., the OSS project team). 
In \SE\ attacks, social engineers are like actors on a stage: they gauge the reaction of a software development team to their performance and adjust it accordingly. Thus, we can use this method to encode  patterns of social behavior that lead to changes in project activity (e.g., dependency management, new functionality), which could open the door to attacker involvement. 
Building upon the discovery of these patterns, we can effectively monitor how specific fragile team dynamics are exploited in other OSS projects. Specifically, we can identify when the exploitation of certain team actions begins, when this exploitation becomes stable, and when there are changes in attacker tactics. 

\subsection{Toward automatic TEMPT prediction}
\label{subsec:attempt}

The amount of supervised training data available to help us predict \SE-based supply-chain attacks may be limited because the literature does not indicate that this type of attack has  been
extensively studied. However, most GitHub projects' metadata
contains a substantial amount of unsupervised data. 
Queries can be formulated to study popular OSS projects on GitHub, and the results can be used to identify generic patterns of team dynamics. Then, these patterns can be used to build a machine learning model using a relatively small amount of training data. 
This technique has been successfully used in natural language processing (NLP), a method that allows learning of temporal word embeddings on unlabeled Wikipedia data \cite{mikolov2017advances}. 

The most suitable evaluation metrics for machine-learning models depend on various factors, such as whether the problem is one of classification or regression. For classification (e.g., predicting whether the project has experienced an \SE\ attack), widely used metrics include \textit{Accuracy}, \textit{Classification Error} ($1$ - Accuracy), \textit{Precision}, \textit{Recall}, and \textit{F-score}. For regression, metrics such as mean absolute error (MAE), minimum sum of absolute error (MSAE), and mean squared error (MSE) are widely employed. 

\section{Related Work}
\label{sec:relatedwork}

\noindent
{\bf Team Practices in OSS Projects:} Today, OSS is more than publicly accessible and available  software. 
Considering the well-established practice of relying upon third-party code dependencies, the software engineering community has studied developer and team practices in  OSS ecosystems. Early studies~\cite{HarsMotivation02,Lakhani03} investigated what motivates developers to take part in OSS projects. Hars and Ou \cite{HarsMotivation02} found  reasons included self-determination, developing human capital, peer recognition, self-marketing, and a developer's personal need to contribute in the  project. Lakhani and Wolf~\cite{Lakhani03} found developers were motivated by the need to contribute, and the opportunity for intellectual stimulation and skill improvement. 

Researchers also recognized  the OSS ecosystems as social networks and investigated how developer and team practices unfold in such an environment.  Ducheneaut ~\cite{Ducheneaut05} was one of the first to describe this social network aspect and the need for new contributors to ``groom'' core members of a project to ensure their contributions were accepted. Bird et al. ~\cite{Bird06} introduced an infrastructure to facilitate the mining of communication information of developers from the Apache HTTP server project  on project archives. Their analysis indicated that the in-degree and the out-degree distribution of the developer social network exhibited typical long-tailed, small-world characteristics and that the level of activity in the source code was a strong indicator of the social status of a developer. However, Guzzi et al.~\cite{Guzzi13} quantitatively and qualitatively analyzed email threads from the development mailing list of the Lucene  OSS project and found that email communication was not the preferred source of team communication.

Tsay et al.~\cite{Tsay14} looked at contributor influence in GitHub via analysis of pull requests.  The analysis compared technical and social measures with the likelihood of contribution acceptance. Pull requests with many comments were less likely to be accepted and acceptance was moderated by the submitter's prior interaction in the project. Well-established projects were more conservative in accepting pull requests.  Trockman et al.~\cite{Trockman18} found a positive correlation between OSS project quality and the use of developer badges (signaling expertise, experience, and commitment). \\

\noindent
{\bf Vulnerability in the Software Supply Chain:} 
Research on protecting registries and OSS project forges is still in its infancy and has focused solely on technical approaches. Zimmermann et al.~\cite{zimmermann2019small} studied the  security threats in the npm ecosystem, describing how  malicious packages were published and how a malicious social engineer became a trusted maintainer. 
The research also described typosquatting (adding a malicious package with a name similar to a legitimate one), and account takeover due to authentication weaknesses. The authors recommend a combination of trusted maintainers and  code vetting to reduce security risks. Duan et al.~\cite{duan2020measuring} introduced  MALOSS, a framework to flag suspicious packages in registries, which combines metadata, and static and dynamic analysis; MALOSS  found several pieces of malware in three main registries: PyPI, npm and RubyGems.  
To the best of our knowledge, no prior work has proposed using fragile team dynamics as predictors of team susceptibility to social engineering attacks that introduce vulnerability in OSS projects.

\section{Conclusion}
\label{sec:conclusion}

As OSS is now critical to the software supply chain,  attacks  can have deep impacts on today's software-reliant systems. The OSS model of productive collaboration among strangers has been its greatest strength but may also prove to be its greatest weakness.  \SE\ attacks against OSS projects are no longer isolated cases and are likely to become a highly exploitable attack avenue, threatening the software security of the entire OSS model. These human-centric attacks leverage common individual and team behaviors (e.g., cognitive heuristics and biases and patterns of behavior), and enable social engineers to apply their deceptive methods across a variety of OSS projects.  While the characteristics of the software may vary, the weaknesses of the humans creating it are often the same.

Our proposed security approach takes a socio-technical, data-driven approach to predict susceptibility of OSS projects to \SE\ attacks in a programming language-, platform-, and vulnerability-agnostic fashion. Automatic identification of  team dynamics could enable teams to self-diagnose fragile behaviors and make changes to strengthen their resistance to \SE\ attacks.  Open and closed-source developers could use a team dynamics assessment to avoid creating risky dependencies on code created or maintained by teams
with fragile dynamics.

This security approach will expose new challenges for the research community.  We anticipate that the study of team dynamics to identify \SE\ attack susceptibility in OSS projects will motivate research in several related areas, including: (1) Support for building ``What if?'' scenarios to expand community understanding of the effects
of team dynamics on the risk of successful \SE\ attacks. Data augmentation
strategies, including simulation and synthetic data generation, could help us to understand the impacts of potential changes in team dynamics. 
(2) Refinement of evaluation criteria, metrics, testing methodologies,
benchmarking, and even methods of formal verification to assess characteristics
of team dynamics. (3) Development of effective proxies of \SE\ susceptibility to generate ground
truths and to enable the design of better learning models. (4) Techniques for improving the robustness of machine
learning models to discern changes in team dynamics as contributors, circumstances,
and team interests  change over time. (5) Development of approaches to detect and mitigate development team members' attempts to influence the team dynamics measurements for their projects. Development teams   may
not appreciate poor ``scores'' of their team dynamics and may seek to ``game the
system'' using manual or automated injections of content (e.g., comments, code
submissions) that mislead our assessment tools. Note that this
last concern does not address social engineers who try to evade detection, but,
rather, OSS developers who object to scrutiny applied to their behavior toward
other team members. Furthermore, our proposed approach may or may not be applicable to closed-source (e.g., corporate) environments.  In such environments, team dynamics ``scores'' may be used to evaluate employee performance. Labor statutes and union contracts may then emerge to control how such information is collected or applied.

Many more research challenges will arise as we study \SE\
attacks against OSS projects. We encourage colleagues in the cybersecurity,
software engineering, social sciences, and other research communities to
contribute their expertise and use their imaginations to address them.

\bibliographystyle{IEEEtran}
\bibliography{main.bib}
%




\end{document}